\documentclass[11pt]{article}
\pdfoutput=1

\usepackage[usenames,dvipsnames]{xcolor}
\usepackage[a4paper]{geometry}
\usepackage[english]{babel}
\usepackage{cite,enumerate,booktabs,float}
\usepackage[affil-it]{authblk}

\usepackage[T1]{fontenc}
\usepackage[utf8]{inputenc}
\usepackage{lmodern}

\usepackage{bm,amsmath,amssymb,tensor,mathtools}
\numberwithin{equation}{section}

\usepackage{tikz}
\usetikzlibrary{cd}

\usepackage{caption}
\usepackage{subcaption}

\usepackage[pdftex]{hyperref}
\definecolor{dark-blue}{rgb}{0.15,0.15,0.4}
\hypersetup{
    colorlinks
    ,linkcolor={dark-blue},
    citecolor={blue}
}

\author{Sam van Leuven\footnote{
	e-mail:
	\href{mailto:S.P.G.vanLeuven@UvA.nl}{S.P.G.vanLeuven@UvA.nl}} }
\author{Gerben Oling\footnote{
	e-mail:
	\href{mailto:G.W.J.Oling@UvA.nl}{G.W.J.Oling@UvA.nl}}
}
\affil{
	Institute for Theoretical Physics,
	University of Amsterdam,\\
	Science Park 904, 1098 XH Amsterdam,
	The Netherlands}
\title{
	Generalized Toda Theory from Six Dimensions\\ and the Conifold
	\vspace{1\baselineskip}
	}
\date{August, 2017}

\newcommand{\RR}{\mathbb{R}}
\newcommand{\CC}{\mathbb{C}}
\newcommand{\ZZ}{\mathbb{Z}}

\newcommand{\vphi}{\varphi}

\newcommand{\OOO}[1]{\mathcal{O}\left(#1\right)}
\newcommand{\LL}{\mathcal{L}}
\DeclareMathOperator{\Tr}{Tr}

\begin{document}

\maketitle

\begin{abstract}
	Recently, a physical derivation of the Alday-Gaiotto-Tachikawa correspondence has been put forward.
	A crucial role is played by the complex Chern-Simons theory arising in the 3d-3d correspondence, whose boundary modes lead to Toda theory on a Riemann surface.
	We explore several features of this derivation and subsequently argue that it can be extended to a generalization of the AGT correspondence.
	The latter involves codimension two defects in six dimensions that wrap the Riemann surface.
  We use a purely geometrical description of these defects and find that the generalized AGT setup can be modeled in a pole region using generalized conifolds.
  Furthermore, we argue that the ordinary conifold clarifies several features of the derivation of the original AGT correspondence.
\end{abstract}

\newpage
\tableofcontents

\section{Introduction}
The Alday-Gaiotto-Tachikawa (AGT) correspondence is a remarkable relation between BPS sectors of four-dimensional supersymmetric gauge theories and two-dimensional non-supersymmetric conformal field theories \cite{alday_liouville_2010,wyllard_a_n-1_2009}.
The correspondence states that $S^4$ partition functions \cite{pestun_localization_2012} of class $\mathcal{S}$ theories of type $A_{N-1}$ \cite{gaiotto_n2_2012} can be expressed as correlation functions in $A_{N-1}$ Toda theory.
In particular, the conformal blocks of the Toda theory were shown to be equivalent to the instanton partition functions, computed in the $\Omega$ background \cite{nekrasov_seiberg-witten_2003}, whereas the three-point functions reproduce the one-loop determinants.

The correspondence can arguably be viewed as the culmination of a long effort towards the understanding of the non-perturbative structure of $\mathcal{N}=2$ Yang-Mills theories \cite{seiberg_monopole_1994,seiberg_monopoles_1994,witten_solutions_1997,nekrasov_seiberg-witten_2003,gaiotto_n2_2012}.
In particular, the systematic construction of the class $\mathcal{S}$ theories provided great insight into the strong coupling limits of these super Yang-Mills theories \cite{gaiotto_n2_2012}.
In this construction, gauge couplings are identified with the complex structure parameters of a Riemann surface and strong-weak dualities are interpreted as a change of `pairs of pants'-decomposition of the Riemann surface.
The AGT correspondence then explicitly brings (non-perturbative) four-dimensional Yang-Mills into the realm of two-dimensional CFT.
This connection is fruitful since the latter class of theories is in general much better understood.
For example, S-duality invariance of the $S^4$ partition function of $\mathcal{N}=2$ $SU(2)$ Yang-Mills with $N_f=4$ corresponds to crossing symmetry of the Liouville four-point function, which was rigorously proven some time ago  \cite{teschner_liouville_1995}.

Increasing the rank, however, the AGT correspondence maps unsolved problems in the gauge theory to other unsolved problems in Toda theory.
For example, the computation of partition functions of non-Lagrangian theories is mapped onto the determination of a general three-point function.
However, the correspondence allows these problems to be phrased in very distinct settings, leading to new insights and progress \cite{mitev_toda_2014,isachenkov_toda_2014}.
Moreover, a complete solution to either problem would kill two birds with one stone. \\

A physical interpretation of the AGT correspondence and its generalizations to higher rank and inclusion of defects seems to rely on a six-dimensional perspective.\footnote{
Relevant references will be given in the main body of the paper.
}
Indeed, the construction of class $\mathcal{S}$ theories already hints at this since it assigns a class $\mathcal{S}$ theory of type $A_{N-1}$ to a punctured Riemann surface $\Sigma$, by compactifying $N$ M5 branes on $\Sigma$ \cite{gaiotto_n2_2012}.
It is precisely this Riemann surface on which the Toda theory lives.
The number of punctures denotes the number of primary insertions in the Toda correlation function.

To be precise, the six-dimensional interpretation of the AGT correspondence is that the supersymmetric partition function of the 6d $(2,0)$ theory $\mathcal{T}$ of type $A_{N-1}$ on $S^4\times \Sigma$ has a four- and two-dimensional incarnation, which are equal.
This is illustrated by the following diagram.
\begin{center}
  \begin{tikzcd}
  & \arrow[ld, black,"\Sigma\to 0" above left] Z_{\mathcal{T}}\left(S^4\times \Sigma\right)\arrow[rd,black,"S^4 \to 0" above right]  & \\
   Z_{\mathcal{S}}\left(S^4\right) & \Longleftrightarrow & Z_{\mathrm{Toda}}\left(\Sigma\right)
  \end{tikzcd}
\end{center}
The arrows denote a supersymmetric zero-mode reduction to the gauge theory $\mathcal{S}$ and Toda theory respectively.
The equivalence of the lower two partition functions is explained through a topological twist performed on $\Sigma$ and the Weyl invariance of $\mathcal{T}$.
These features enable us to send the size of either manifold to zero without affecting the value of the partition function, as long as we restrict to the supersymmetric sector.
However, the lack of a Lagrangian description of $\mathcal{T}$ blocks a straightforward implementation of this strategy.

Over the past few years, many different approaches have been taken to overcome this difficulty.
See for an incomplete list of references \cite{alday_liouville/toda_2010,nekrasov_omega_2010,mironov_direct_2011,dijkgraaf_toda_2009,aganagic-triality-2014,yagi_compactification_2012,tan_m-theoretic_2013,vartanov_supersymmetric_2013,beem-w-2015}.
A constructive derivation of the correspondence is desirable as it could provide an idea of the scope of AGT-like correspondences between supersymmetric sectors of gauge theories and exactly solvable models.
Moreover, due to its six-dimensional origin, such a derivation may also shed light on the worldvolume theory of multiple M5 branes.\\

In this paper we will build on a recent derivation by Córdova and Jafferis \cite{cordova-toda-2016}.
Using the relation between the type $A_{N-1}$ 6d $(2,0)$ theory on a circle and five-dimensional $\mathcal{N}=2$ $SU(N)$ Yang-Mills theory \cite{seiberg_five_1996,seiberg_notes_1998,douglas_d5_2011}, one performs a Kaluza-Klein reduction on $S^4$ to obtain $A_{N-1}$ Toda theory on a Riemann surface $\Sigma$.
The Toda fields are understood as boundary fluctuations of $SL(N,\CC)$ Chern-Simons theory on a manifold with asymptotically hyperbolic boundary.
This is understood in the following way.
Near the boundary, the Chern-Simons connection satisfies the boundary conditions
\begin{equation}\label{eq:cs-connection-bcs}
	\mathcal{A} \to
		\frac{d\sigma}{\sigma} H + \frac{du}{\sigma} T_+
			+ \OOO{\sigma^0}.
\end{equation}
Here, $H$ is an element of the Cartan of $sl_N$, which sits together with a raising operator $T_+$ in an $sl_2\subset sl_N$ subalgebra.
In a type IIA frame, these boundary conditions arise from a Nahm pole on the scalars of D4 branes ending on D6 branes \cite{diaconescu_d-branes_1997,gaiotto_supersymmetric_2009}.

Boundary conditions such as \eqref{eq:cs-connection-bcs} are well known to provide a reduction of the $\widehat{sl_N}$ WZW theory induced by Chern-Simons on the boundary of asymptotically hyperbolic space, see for example \cite{bais_covariantly_1991,de_boer_relation_1994,de_boer_thermodynamics_2014}.
For the principal $sl_2$ embedding found in \cite{cordova-toda-2016}, such constraints give Toda theory \cite{forgacs-liouville-1989}.
Consequently, one of the building blocks in establishing the AGT correspondence is obtained.

However, the residual symmetries of the constrained WZW theory strongly depend on the embedding of $sl_2$ into $sl_N$.
For example, for $N=3$, the reduced boundary theory has $\mathcal{W}_3$ symmetry if the embedding is the principal one, but it has Polyakov-Bershadsky $\mathcal{W}_3^{(2)}$ symmetries for the diagonal embedding.
More generally, $sl_2$ embeddings into $sl_N$ are labeled by the integer partitions $\lambda$ of $N$.
Each choice leads to a reduced boundary theory with different symmetries, which we will denote by $\mathcal{W}_\lambda$.
These generalized Toda theories play a role in extensions of the AGT correspondence.\\

In \cite{alday_affine_2010} a relation was proposed between instanton partition functions of $\mathcal{N}=2$ $SU(2)$ quiver gauge theories with an insertion of a surface operator, which arises from a codimension two defect in the 6d theory, and conformal blocks of $\widehat{sl_2}$ WZW theories.
This was generalized in \cite{kozcaz_affine_2011} to a relation between $SU(N)$ gauge theories and $\widehat{sl_N}$ WZW theories.
These cases dealt with the so-called full surface operators.

It was conjectured in \cite{wyllard_w-algebras_2011} that the $SU(N)$ instanton partition functions with more general surface operators, labeled by a partition $\lambda$ of $N$, would be equivalent to the conformal blocks of theories with $\mathcal{W}_{\lambda}$ symmetry.
The standard AGT and full surface operator setup are now special cases of this more general setup, corresponding to the partitions $N=N$ and $N=1+\ldots +1$ respectively.
The $\mathcal{W}_{\lambda}$ algebra, which is also labeled by a partition of $N$, is obtained by quantum Drinfeld-Sokolov reduction of $\widehat{sl_N}$.
An explicit check was performed for the Polyakov-Bershadsky algebra $\mathcal{W}_3^{(2)}$, whose conformal blocks were shown to agree with instanton partition functions in the presence of a simple surface defect, with partition $3=2+1$.
Further checks of the proposal have appeared in \cite{wyllard_instanton_2011,tachikawa-walgebras-2011}.

Then, based on mathematical results in instanton moduli spaces, it was realized in \cite{kanno-instanton-2011} that the instanton partition function in the presence of a general surface operator on $\CC^2$ could be conveniently computed as an ordinary instanton partition function on $\CC/\ZZ_m\times \CC$, where $m$ corresponds to the maximum number of parts of the partition $\lambda$.
This technique was further used in \cite{nawata_givental_2014} to compute the $S^4$ partition functions of $\mathcal{N}=2^*$ $SU(N)$ theories in the presence of a full surface operator, and was shown in the case of $SU(2)$ to reproduce the full $\widehat{sl_2}$ WZW correlation function.
For $SU(N)$ results were obtained as well, but could not be compared due to lack of results on the WZW side.

In the following, we will denote the generalized Toda theory resulting from an $sl_N$ reduction with partition $\lambda$ by $\text{Toda}_\lambda$.
The corresponding generalized AGT correspondence will be referred to as the $\text{AGT}_\lambda$ correspondence.
\\

In the present paper, we propose a setup to derive these $\text{AGT}_\lambda$ correspondences using the path laid out by Córdova and Jafferis.
This approach is very natural for the problem at hand, since the general quantum Drinfeld-Sokolov reduction of $\widehat{sl_N}$ can be understood from a Chern-Simons perspective as well, by imposing the boundary conditions \eqref{eq:cs-connection-bcs} for a general $sl_2\subset sl_N$ embedding.
Therefore, we wish to show that upon including the appropriate codimension two defects in the six-dimensional setup, one finds these more general boundary conditions.
Along the way, we will also be able to clarify some aspects of the analysis in the original paper \cite{cordova-toda-2016}.

\subsection{Overview and summary of results}
\label{ssec:overview-summary}
Since the story is rather intricate and hinges on some important assumptions, we will briefly sketch the main logic and possible pitfalls of our arguments here.

The original derivation, which we review in section \ref{ssec:cj-review}, connects the 4d-2d correspondence to the 3d-3d correspondence through a Weyl rescaling.
One of the main virtues of this connection is that a full supergravity background was already derived in \cite{cordova_complex_2013} for the 3d-3d correspondence, which can then be put to use in the 4d-2d setting.
The three-manifold $M_3$ on which the resulting Chern-Simons theory lives has nontrivial boundary.
With specific boundary conditions, its boundary excitations lead to Toda theory.

These boundary conditions manifest themselves in a IIA frame in the form of a Nahm pole on the worldvolume scalars of a D4 brane ending on a D6 brane.
The original derivation attributes the Nahm pole to the D6 branes that are also related to a non-zero Chern-Simons level.
We point out that the Nahm pole should instead be attributed to a distinct set of branes, which we refer to as D6' branes.
The original branes will always be referred to as D6 branes, and will still be related to the Chern-Simons coupling.\\

A crucial element in the original derivation is that the Nahm pole on the scalars transforms under Weyl rescaling to the relevant Drinfeld-Sokolov boundary condition on the Chern-Simons connection.
It is argued that this boundary condition is a natural way to combine Nahm data into a flat connection, but the Drinfeld-Sokolov form is not the unique combination that achieves this.
However, we have not been able to obtain a better understanding of this point and our construction still relies on this assumption.
We expect that carefully examining the Weyl rescaling of the full supergravity background and the corresponding worldvolume supersymmetry equations should allow one to translate the Nahm pole arising in the 4d-2d frame to the Drinfeld-Sokolov boundary condition in the 3d-3d frame.
However, a direct implementation of this procedure is ruled out by the lack of a Lagrangian description of multiple M5 branes.

The Weyl rescaling of the full supergravity background should also allow one to further explain the claim in \cite{cordova-toda-2016} that the Killing spinors as obtained in \cite{cordova_complex_2013} for the 3d-3d background become the usual 4d Killing spinors of \cite{pestun_localization_2012,hama-seibergwitten-2012} after Weyl rescaling and an R-gauge transformation.
This argument is not completely satisfactory, since the spinors in the 3d-3d frame are related to a squashed sphere geometry that preserves an $SU(2)\times U(1)$ isometry, whereas the Killing spinors in \cite{hama-seibergwitten-2012} are related to a squashed sphere with $U(1)\times U(1)$ isometry.
We note that this slight discrepancy may in fact be immaterial at the level of partition functions, as was indeed originally found in \cite{imamura_n2_2012} in the context of 3d partition functions and properly understood in \cite{closset_geometry_2014}.
\\

The uplift to M-theory of the setup we propose leads to M5 branes on a holomorphic divisor in a generalized conifold, which we discuss in section \ref{sec:defects-conifold}.
Here, we crucially use the orbifold description of codimension two defects that was advocated in \cite{tachikawa-walgebras-2011,kanno-instanton-2011}.\footnote{
  The gravity duals of class $\mathcal{S}$ theories similarly treat such codimension two defects geometrically \cite{gaiotto-gravity-2012}.
}
This enables us to treat the defects purely geometrically, so that we do not have to worry about coupling the worldvolume theory to additional degrees of freedom on the defect.

We propose to use the conifold geometry as an approximation to the pole region of a full supergravity background that would be needed to account for a defect in a squashed $S^4$ background.\footnote{See also \cite{gukov_equivariant_2015,gukov_equivariant_2016} for a like-minded approach to the 3d-3d correspondence.}
Although this approximation suffices for our purposes, it comes with a particular value of the squashing parameter that leads to a curvature singularity corresponding to the conifold point.
In principle, such a singularity could couple to the M5 worldvolume theory.
It would therefore be very interesting to obtain a class of supergravity backgrounds for arbitrary parameter values where this singularity can be avoided.

The radial slices of the divisor of the generalized conifold have a $U(1)\times U(1)$ isometry.
Furthermore, it supports two supercharges, in agreement with the four-dimensional $\Omega$ background.
In the special case where only a single D6' brane is present, corresponding to a trivial surface operator, the isometry enhances to $SU(2)\times U(1)$, but still only two supercharges are present.
This may seem strange, since one expects a nontrivial surface operator to break part of the supersymmetries.
However, placing a surface operator on a fully squashed $S^4$ does not break any additional isometries, hence the number of preserved supercharges on a fully squashed background is the same with or without a surface operator.
\\

An important assumption in our derivation is that the connection to the 3d-3d correspondence still stands.
Even though additional defects are present we claim that these only manifest themselves in the boundary conditions of the Chern-Simons theory.
Since these defects are located at the asymptotic boundary of $M_3$, we believe that this claim is justified.

Finally, it is known that at $k=1$ the Hilbert spaces of $SL(N,\CC)$ and $SL(N,\RR)$ Chern-Simons theories agree \cite{dimofte-complex-2014}.
Therefore, at $k=1$ the reduction to generalized real Toda theories proceeds as usual.
For higher $k$, one obtains complex $\text{Toda}_\lambda$ theories.
In the principal case, the original derivation puts forward a duality between complex Toda and real paraToda with a decoupled coset.
It would be interesting to formulate a similar correspondence for complex $\text{Toda}_\lambda$ theories.

\section{Review}\label{sec:review}
In this section we review the derivation by Córdova and Jafferis of both the 3d-3d and AGT correspondence \cite{cordova_complex_2013,cordova-toda-2016}.
Subsequently, we give an overview of the relation between Chern-Simons theory and Wess-Zumino-Witten models and their Drinfeld-Sokolov reduction to $\text{Toda}_\lambda$ theories.

\subsection{Principal Toda theory from six dimensions}
\label{ssec:cj-review}
\begin{figure}
\captionsetup{subrefformat=parens}
	\begin{subfigure}[]{\linewidth}
		\centering
		\includegraphics
			[width=.8\textwidth]
			{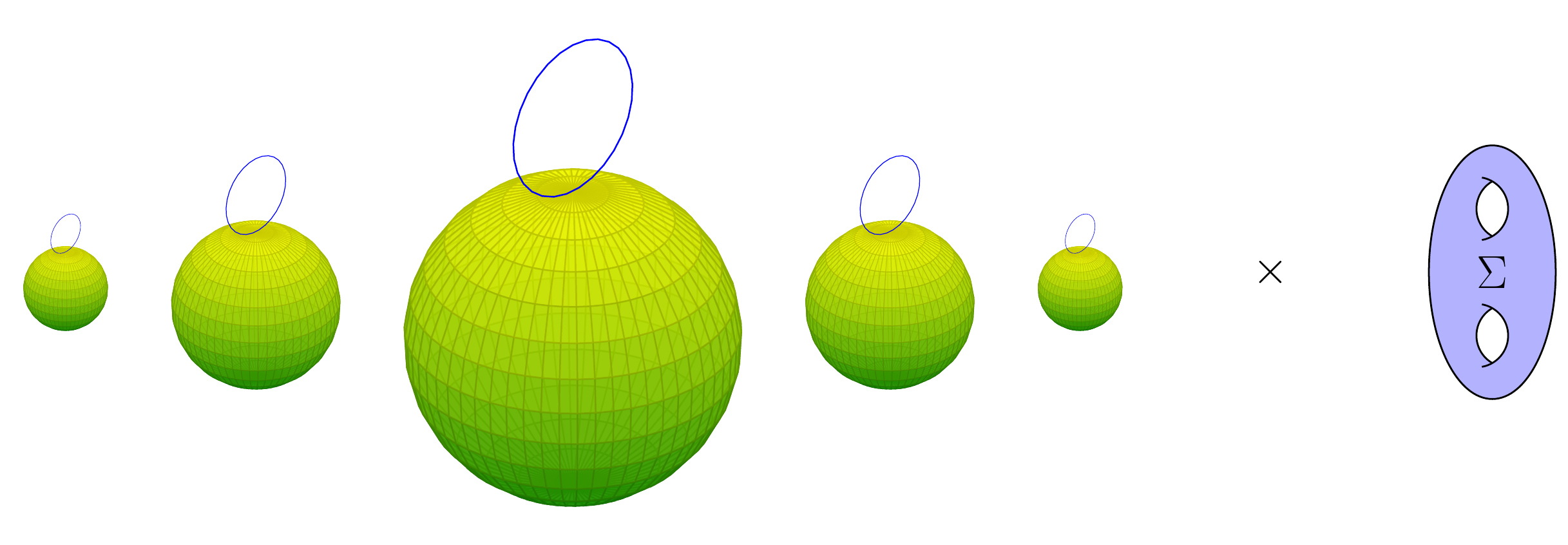}
		\caption{
			$S_{\ell}^4/\ZZ_k\times \Sigma$
		}
		\label{fig:4d2d-frame}
	\end{subfigure}
	\par\bigskip
	\begin{subfigure}[]{\linewidth}
		\centering
		\includegraphics
			[width=.8\textwidth]
			{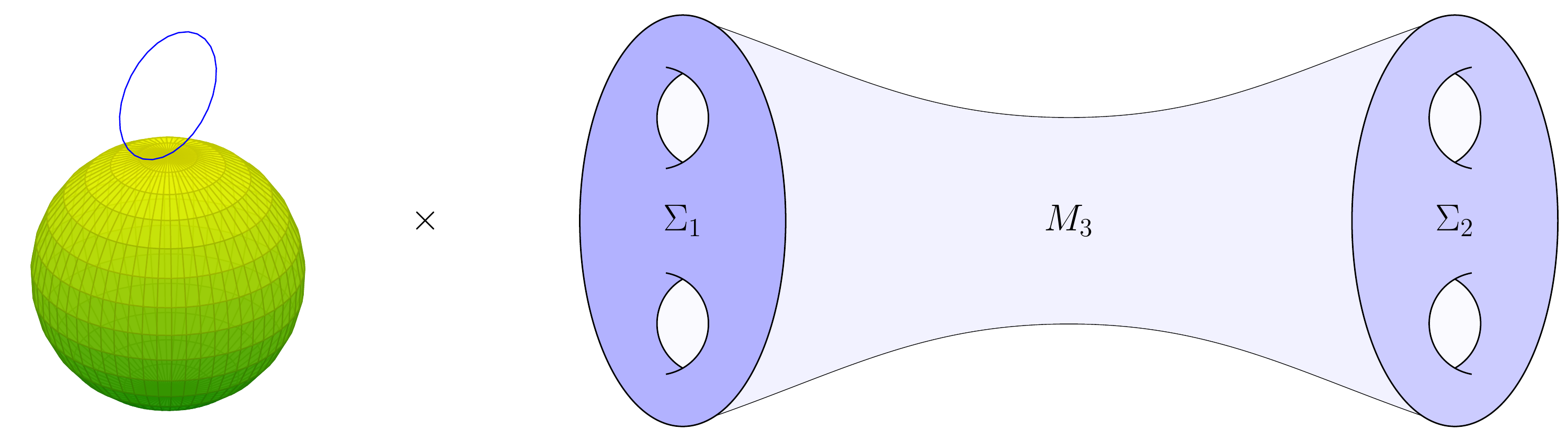}
		\caption{
			$S_{\ell}^3/\ZZ_k \times M_3$
		}
		\label{fig:3d3d-frame}
	\end{subfigure}
	\caption{The geometries associated to \subref{fig:4d2d-frame} the 4d-2d frame and \subref{fig:3d3d-frame} the 3d-3d frame. The Hopf fiber of the $S^3$ is indicated in blue.}
	\label{fig:4d2d-3d3d-frame}
\end{figure}

Consider the 6d $(2,0)$ CFT of type $A_{N-1}$ on two geometries which are related by a Weyl transformation \cite{cordova-toda-2016}
\begin{equation}\label{eq:weyl-transf}
	S_{\ell}^4/\ZZ_k\times \Sigma
		\; \xLeftrightarrow{\text{Weyl}} \;
	S_{\ell}^3/\ZZ_k \times M_3.
\end{equation}
We think of the $S^4/\ZZ_k$ as the Lens space $S^3/\ZZ_k$ fibered over an interval, shrinking to zero size at the endpoints.
The three-dimensional manifold $M_3$ is a warped product of a Riemann surface $\Sigma$ and $\RR$ and $\ell$ is a squashing parameter which controls the ratio between the Hopf fiber and base radius of the $S^3$.
We will refer to these geometries as the 4d-2d and 3d-3d geometries respectively.
See figure \ref{fig:4d2d-3d3d-frame} for an illustration.

In older work \cite{cordova-fivedimensional-2013} it was shown how to couple 5d $\mathcal{N}=2$ SYM to 5d $\mathcal{N}=2$ off-shell supergravity.
Using the equivalence between the $A_{N-1}$ $(2,0)$ theory on a circle and 5d $\mathcal{N}=2$ $SU(N)$ Yang-Mills theory \cite{seiberg_five_1996,seiberg_notes_1998,douglas_d5_2011}, these general results allow one to preserve four supercharges from the $(2,0)$ theory on the geometry \cite{cordova_complex_2013}
\begin{equation}\label{eq:6dgeom-embedded-in-11d}
S_{\ell}^3/\ZZ_k\times M_3\subset S_{\ell}^3/\ZZ_k\times T^*M_3\times \RR^2.
\end{equation}
In the original derivation, the $(2,0)$ theory is reduced on the Hopf fiber.
This translates in 5d to a flux for the graviphoton, which is compatible with the 5d supergravity background.
For general squashing, it is required to turn on all bosonic fields in the off-shell supergravity multiplet.
The resulting background allows for a supersymmetric zero mode reduction on the $S^3$ which gives rise to $SL(N,\CC)$ Chern-Simons on $M_3$ with coupling $q=k+i\sqrt{\ell^2-1}$.\footnote{
This provides a derivation of the 3d-3d correspondence as formulated in \cite{dimofte_vortex_2010,dimofte_gauge_2011,dimofte_3-manifolds_2011}.}

The complex Chern-Simons coupling consists of an integer $k$ and a continuous parameter $\ell$.
The former arises from the graviphoton flux that couples to the D4 gauge fields through the 5d Chern-Simons coupling
\begin{equation}
  \label{eq:graviphoton-flux-cs-couping}
  \frac{1}{8\pi^2} \int_{S^2\times M_3}
    \Tr\left( C \wedge F \wedge F\right)
  \implies \frac{k}{4\pi} \int_{M_3}
    \Tr\left( A\wedge dA + \frac{2}{3} A \wedge A\wedge A \right).
\end{equation}
The continuous parameter $\ell$ arises from the squashing parameter of the three-sphere.

A salient detail of the reduction is that the fermions of the $(2,0)$ theory come to be interpreted as Faddeev-Popov ghosts for the gauge fixing of the non-compact part of the gauge algebra
$$sl(N,\CC)\cong su(N)\oplus i \,su(N).$$
This provides a concrete explanation for the puzzle that the supersymmetric reduction of 5d \textit{supersymmetric} Yang-Mills with \textit{compact} gauge group $SU(N)$ becomes a \textit{non-supersymmetric} Chern-Simons theory with \textit{non-compact} gauge group.
The ghost and gauge fixing terms in the effective action are subleading in $R_{S^3}$, so that in the far IR the gauge fixing is undone and the final result for the effective theory on $M_3$ is the full $SL(N,\CC)$ Chern-Simons theory.
The complex connection
$$\mathcal{A}=A+iX$$
is built out of the original Yang-Mills connection together with three of the five worldvolume scalars $X_i$.
The latter combine into a one-form on $M_3$ due to the topological twist.
We denote the other two scalars by $Y_a$.
They correspond to movement in the remaining $\RR^2$ directions of \eqref{eq:6dgeom-embedded-in-11d}.

We now return to the particular $M_3$ that arises from the Weyl rescaling of the 4d-2d background.
Note that it has a nontrivial boundary consisting of two components $\Sigma\cup\Sigma$.
So we need to specify boundary conditions, which ultimately lead to non-chiral complex Toda theory on $\Sigma$, as we will review in section \ref{ssec:dsreduction}.

To understand what type of boundary conditions have to be imposed, let us first look at the following table that summarizes the 4d-2d setup:
\begin{table}[H]
\centering
\begin{tabular}{lccccccccccc}
\toprule
& \multicolumn{4}{c}{$S^4/\ZZ_k$}&   \multicolumn{2}{c}{$\Sigma$}& \multicolumn{3}{c}{$\RR^3$}& \multicolumn{2}{c}{$\RR^2$}\\
\midrule
  & 0               & 1               & 2               & 3  & 4  & 5  & 6  & 7 & 8  & 9 &10  \\
\midrule
$N$ M5      & x              &x              & x              & x & x&x && & &&  \\
\bottomrule
\end{tabular}
\caption{M-theory background relevant for the AGT correspondence.}
\label{tab:m}
\end{table}
\noindent The theory is topologically twisted along $\Sigma$.
An $\RR^2\subset \RR^3$ provides the fibers of its cotangent bundle $T^*\Sigma$.
In the 3d-3d frame the entire $\RR^3$ is used for the topological twist on $M_3$.

The setup is reduced on the Hopf fiber of the $S^3/\ZZ_k\subset S^4/\ZZ_k$.
Equivalently, thinking of the $S^4$ as two $k$-centered Taub-NUTs glued along their asymptotic boundary, one reduces on the Taub-NUT circle fiber.
It is well known that the M-theory reduction on the circle fiber of a multi-Taub-NUT yields D6 branes at the Taub-NUT centers.
The IIA setup\footnote{
Note that the resulting three-sphere has curvature singularities at the poles even for $k=1$.}
is then given by Table \ref{tab:iia}.
\begin{table}[H]
\centering
\begin{tabular}{lccccccccccc}
\toprule
& \multicolumn{3}{c}{$S^3$}&   \multicolumn{2}{c}{$\Sigma$}& \multicolumn{3}{c}{$\RR^3$}& \multicolumn{2}{c}{$\RR^2$}\\
\midrule
                 & 1               & 2               & 3  & 4  & 5  & 6  & 7 & 8  & 9 &10  \\
\midrule
$N$ D4                    &x              & x              & x & x&x && & &&  \\
$k$ D6            &    &              &  & x&x &x&x &x &x&x\\
$k$ $\overline{\mathrm{D6}}$         &        &              &  & x&x &x&x &x &x&x\\
\bottomrule
\end{tabular}
\caption{Reduction of M-theory background to type IIA.}
\label{tab:iia}
\end{table}
\noindent The D6 and $\overline{\mathrm{D6}}$ branes sit at the north and south pole of the $S^3$ respectively and the D4 branes end on them.
The boundary conditions on the D4 worldvolume fields are then claimed to be similar to those studied in \cite{gaiotto_supersymmetric_2009} for D3 branes ending on D5 branes.
That would imply that the D4 gauge field satisfies Dirichlet boundary conditions, while the triplet of scalars $X_i$ satisfy the Nahm pole boundary conditions
\begin{equation}\label{eq:nahm-pole-scalars}
X_i\to \frac{T_i}{\sigma}.
\end{equation}
Here, the $T_i$ constitute an $N$ dimensional representation of $su(2)$ and $\sigma$ parametrizes the interval over which the $S^3/\ZZ_k$ is fibered.
This can be understood by thinking of the $N$ D4 branes as comprising a charge $N$ monopole on the D6 worldvolume.
Indeed, the Nahm pole boundary conditions were originally discovered in a similar context \cite{diaconescu_d-branes_1997}.\\

We want to pause here for a moment to note that it is not quite clear why the present setup is related to the analyses of \cite{diaconescu_d-branes_1997,gaiotto_supersymmetric_2009}.
The latter deal with (the T-dual of) a D4-D6 brane system with different codimensions, as described in table \ref{tab:nahmpole}.
\begin{table}[H]
\centering
\begin{tabular}{lccccccccccc}
\toprule
&$\RR$ & \multicolumn{2}{c}{$\RR^2$}&   \multicolumn{2}{c}{$\Sigma$}& \multicolumn{3}{c}{$\RR^3$}& \multicolumn{2}{c}{$\RR^2$}\\
\midrule
                 & 1               & 2               & 3  & 4  & 5  & 6  & 7 & 8  & 9 &10  \\
\midrule
$N$ D4                    &$\vdash$              & x              & x & x&x && & &&  \\
$k$ D6            &    &      x        & x & x&x &x&x &x &&\\
\bottomrule
\end{tabular}
\caption{Type IIA setup in which Nahm poles arise as boundary conditions on the $X_i$ triplet of D4 scalars.}
\label{tab:nahmpole}
\end{table}
\noindent Here, the $\vdash$ denotes the fact that the D4 branes end on the D6 branes.
The D6 branes in table \ref{tab:iia} are instead similar to the ones studied in \cite{dijkgraaf_supersymmetric_2008} (see also \cite{itzhaki_i-brane_2006}).
The corresponding orbifold singularities in the 4d-2d frame reduce to a graviphoton flux in the 3d-3d frame, which is responsible for the Chern-Simons coupling through \eqref{eq:graviphoton-flux-cs-couping}.
However, they cannot give rise to a Nahm pole.
In section \ref{sec:toda-from-conifold}, we propose an alternative perspective that simultaneously allows for a non-zero Chern-Simons coupling \textit{and} correct codimensions between D4 and D6 branes for a Nahm pole to arise.\\

Leaving these comments aside for the moment, we must understand precisely what a Nahm pole in the topologically twisted scalars $X_i$ would translate to in the 3d-3d picture.
As remarked in section \ref{ssec:overview-summary}, transforming the supersymmetry equations that lead to a Nahm pole under the Weyl transformation is a dificult problem.
However, we know that the resulting connection $\mathcal{A}=A+iX$ will have to be flat.
Furthermore, we expect that the leading behavior of $\mathcal{A}$ towards the boundary should still be fixed.

As we will review in the following section, the relation between Chern-Simons theory and Wess-Zumino-Witten models requires $\mathcal{A}$ to be chiral on the boundary.\footnote{
Nonchiral boundary conditions lead to reduced theories with nonzero chemical potentials \cite{de_boer_boundary_2014,de_boer_thermodynamics_2014}.
It would be interesting to see if they have a role to play in further generalizations of the AGT correspondence.
}
If $(z,\bar{z})$ denote (anti)holomorphic coordinates on $\Sigma$, we should demand that $\mathcal{A}_{\bar{z}}$ vanishes.
Thus, a natural equivalent of the boundary conditions \eqref{eq:nahm-pole-scalars} would be
\begin{equation}\label{eq:nahm-pole-connection}
	\mathcal{A}
		= \LL_0 \frac{d\sigma}{\sigma} + \LL_+\frac{dz}{\sigma}
			+ \OOO{\sigma^0}.
\end{equation}
This is a flat connection.
We have defined the $sl_2$ generators
	$\LL_0 = iT_1$ and
	$\LL_\pm = T_2 \mp i T_3$.
They satisfy the standard commutation relations
\begin{equation}
	[\LL_a, \LL_b] = (a-b) \LL_{a+b}.
\end{equation}
These boundary conditions are precisely the ones that correspond to the reduction of the boundary $\widehat{sl_N}$ algebra to the $\mathcal{W}_N$ algebra.
The antiholomorphic connection of the complex Chern-Simons theory behaves in the same way.
Adding the contributions from the two components of $\partial M_3$ then gives rise to a full (non-chiral) \emph{complex} Toda theory.
It would be interesting to directly verify the transformation of the Nahm pole \eqref{eq:nahm-pole-scalars} to the connection boundary condition \eqref{eq:nahm-pole-connection} under the Weyl transformation, as we already pointed out in section \ref{ssec:overview-summary}.

\subsection{Partitions of \texorpdfstring{$N$}{N} and Drinfeld-Sokolov reduction}\label{ssec:dsreduction}
On a three-dimensional manifold $M_3$ with boundary, Chern-Simons theory with gauge algebra $sl_N$ induces an $\widehat{sl_N}$ Wess-Zumino-Witten model on $\partial M_3$.
Boundary conditions on the connection such as those we encountered in \eqref{eq:nahm-pole-connection} translate to constraints in the WZW model.
Many of the results we discuss are well known in the literature on WZW models and three-dimensional gravity, see \cite{feher-hamiltonian-1992,bouwknegt-wsymmetry-1993,perez-brief-2014,donnay-asymptotic-2016} for reviews.
We simply wish to point out how they can be used in deriving the $\text{AGT}_\lambda$ correspondence.

We first recall how one obtains a (Brown-Henneaux) Virasoro algebra in the $sl_2$ case \cite{banados-global-1995,coussaert-asymptotic-1995,banados-threedimensional-1999}.
Here we restrict to the holomorphic sector of the complex Chern-Simons theory.
The following holds similarly for the antiholomorphic sector.
The variation of the Chern-Simons action is
\begin{equation}
	\delta S_{CS}
		= \frac{k}{2\pi} \int_{M_3} \Tr[\delta \mathcal{A} \wedge \mathcal{F}]
			+ \frac{k}{4\pi} \int_{\partial M_3}
				\Tr[\mathcal{A}\wedge \delta \mathcal{A}].
\end{equation}
The bulk term would lead us to identify the vanishing of the curvature $\mathcal{F}= d\mathcal{A} + \mathcal{A}\wedge \mathcal{A}$ as the equations of motion.
However, this is not justified unless the boundary term vanishes.
It is most commonly dealt with by requiring one of the boundary components to vanish.
Using coordinates $(z,\bar{z})$ on $\Sigma$, one can set
\begin{equation}\label{eq:cs-chirality-condition}
	\mathcal{A}_{\bar{z}} = 0
		\quad\text{on } \partial M_3.
\end{equation}
With these boundary conditions, Chern-Simons theory describes a Wess-Zumino-Witten model on $\partial M$.
In particular, we can use the bulk gauge freedom to fix the radial component to be
\begin{equation}\label{eq:cs-sl2-radial-cpt-fixed}
	A_\rho = \LL_0 \in sl_2.
\end{equation}
The most general flat connection satisfying \eqref{eq:cs-chirality-condition} and \eqref{eq:cs-sl2-radial-cpt-fixed} is then
\begin{equation}\label{eq:ds-standard-connection-form}
	\begin{split}
	\mathcal{A}
		&= \LL_0 d\rho + e^{\rho} J^+(z)\LL_+ dz
			+ J^0(z)\LL_0 dz + e^{-\rho} J^-(z)\LL_- dz \\
		&= e^{-\rho \LL_0} \left(
				d + J^a(z) \LL_a dz
			\right) e^{\rho \LL_0}.
	\end{split}
\end{equation}
The remaining chiral degrees of freedom $J^a(z)$ are the currents of the chiral Wess-Zumino-Witten model.
One can consider the reduction of this model using certain constraints.
In particular, using the radial coordinate $e^{-\rho} = 1/\sigma$, the leading order of the transformed Nahm pole boundary conditions \eqref{eq:nahm-pole-connection} can be written as
\begin{equation}\label{eq:nahm-pole-connection-for-ds}
	\mathcal{A}_0
	 	= e^{-\rho \LL_0} \left(
				d + \LL_+ dz
			\right) e^{\rho \LL_0}
		= \LL_0 d\rho + e^{\rho} \LL_+ dz.
\end{equation}
Comparing this to the WZW current components in \eqref{eq:ds-standard-connection-form} up to leading order in $\rho$ leads to a first class constraint $J^+ \equiv 1$.
We can use the resulting gauge symmetry to fix $J^0 \equiv 0$, leading to a second class set of constraints.
The reduced on-shell phase space consists of
\begin{equation}
	\mathcal{A} = e^{- \rho \LL_0} \left(
				d + \LL_+ dz + J^-(z)dz
			\right)
			e^{\rho \LL_0}.
\end{equation}
Its residual symmetries form a Virasoro algebra with current $T(z)=J^-(z)$ and central charge $c = 6k$.\footnote{
From the perspective of three-dimensional Einstein gravity, which can be described by two chiral $sl_2$ Chern-Simons actions with $k=l/4G_N$, the reference connection $\mathcal{A}_0$ is empty $AdS_3$.
The reduced symmetry algebra is a chiral half of the Brown-Henneaux asymptotic Virasoro symmetries.
Constraining the leading-order radial falloff corresponds to imposing Dirichlet constraints on the boundary metric of asymptotically $AdS_3$ geometries.
}
On the level of the action, the reduction outlined above produces Liouville theory from the $\widehat{sl_2}$ WZW model.\\

Now let us return to $sl_N$, where the corresponding situation has been studied in the context of current algebras
\cite{forgacs-liouville-1989,feher-hamiltonian-1992,bais_covariantly_1991,de_boer_relation_1994} and higher spin gravity
\cite{campoleoni_asymptotic_2010,campoleoni_asymptotic_2011,de_boer_boundary_2014,de_boer_thermodynamics_2014}.
Again, we can impose chiral boundary conditions \eqref{eq:cs-chirality-condition} and gauge fix the radial component as in \eqref{eq:cs-sl2-radial-cpt-fixed},
\begin{equation*}
	\left.\mathcal{A}_{\bar{z}}\right|_{\partial M_3} = 0, \quad
	\mathcal{A}_\rho = \LL_0 \in sl_N.
\end{equation*}
We denote the $sl_N$ generators by $T_a$.
The chiral connection of \eqref{eq:ds-standard-connection-form} describing the WZW model becomes
\begin{equation}\label{eq:ds-standard-slN-connection-form}
	\mathcal{A} = e^{-\rho \LL_0} \left(
				d + J^a(z) T_a dz
			\right) e^{\rho \LL_0}.
\end{equation}
Its asymptotic behavior is constrained by the leading order behavior of the Weyl transformed Nahm pole,
\begin{equation*}
	\mathcal{A}_0
		= e^{-\rho \LL_0} \left(
			d + \LL^+ dz
		\right) e^{\rho \LL_0}
		= \LL_0 d\rho + e^{\rho} \LL_+ dz.
\end{equation*}
The latter dictates a particular choice of $sl_2\subset sl_N$ embedding through the generators $\{\LL_0, \LL_+\}$ appearing in it.\footnote{
From the point of view of three-dimensional gravity, this choice of embedding corresponds to choosing an Einstein sector within higher spin gravity.
}
It is therefore useful to organize the $sl_N$ basis in multiplets of the $sl_2$ subalgebra corresponding to $\LL_a$.

In particular, the radial falloff of a current component $J^a$ in the connection is then determined by the weight of the corresponding generator $T^a$ under $\LL_0$.
To be precise, if $[\LL_0, T_a] = w_{(a)} T_a$, we see that the $T_a$ component of $\mathcal{A}_z$ is
\begin{equation}
	\left.\mathcal{A}_{z}\right|_{T_a} T_a
		= J^a(z)\, e^{-\rho \LL_0} \,T_a \,e^{\rho\LL_0}
		= e^{ - w_{(a)} \rho} \,J^a(z)\,T_a.
\end{equation}
Thus, if we fix the non-normalizable part of the connection in terms of $\mathcal{A}_0$,
\begin{equation}
	\mathcal{A} - \mathcal{A}_0 \equiv \OOO{1} \quad \text{as}\quad \rho\to \infty,
\end{equation}
we constrain all the current components $J^a$ of $w_{(a)}<0$ generators,
\begin{equation}\label{eq:ds-current-constraints}
	J^{\LL_+} \equiv 1 \quad \text{and} \quad
		J^{T_a} \equiv 0 \quad \text{ for all other negative weight generators } T_a.
\end{equation}
These constraints generate additional gauge freedom, which can be used to fix all but the highest weight currents of each multiplet to zero.
This brings us to what is usually known as highest-weight or Drinfeld-Sokolov gauge, with a single current for each $sl_2$ multiplet.\\

We shortly review $sl_2\subset sl_N$ embeddings and the corresponding multiplet structure.
The possible decompositions of the $sl_N$ fundamental representation are labeled by partitions $\lambda$ of $N$,
\begin{equation}\label{eq:slN-fundamental-decomposition}
	\textbf{N}_N =
		\bigoplus_{k=1}^N n_\text{k} \textbf{k}_2
				\qquad\longleftrightarrow\qquad
	\lambda:\quad N = \sum n_\text{k}	.
\end{equation}
Here, we use $\textbf{k}_M$ to denote a k-dimensional fundamental representation of $sl_M$.
We are interested in the multiplet structure of $sl_N$ under the adjoint action of its $sl_2$ subalgebra.
Through the corresponding decomposition $N=\sum n_\text{k}$ in \eqref{eq:slN-fundamental-decomposition}, the choice of partition $\lambda$ determines the number of $sl_2$ multiplets in the adjoint representation of $sl_N$.
For example, if $N=3$ we can choose $3=3$, $3=2+1$ or $3=1+1+1$, corresponding to
\begin{alignat}{2}
	\bm{3}_3 &= \bm{3}_2
			\quad &&\implies \quad
		\bm{3}_3 \otimes \bm{\bar{3}}_3 - \bm{1}_2
			= \bm{5}_2 \oplus \bm{3}_2 , \\
	\bm{3}_3 &= \bm{2}_2 \oplus \bm{1}_2
			\quad &&\implies \quad
		\bm{3}_3 \otimes \bm{\bar{3}}_3 - \bm{1}_2
			= \bm{3}_2 \oplus 2 \, \bm{2}_2 \oplus \bm{1}_2, \\
	\bm{3}_3 &= \bm{1}_2 \oplus \bm{1}_2 \oplus \bm{1}_2
			\quad &&\implies \quad
		\bm{3}_3 \otimes \bm{\bar{3}}_3 - \bm{1}_2
			= 8 \, \bm{1}_2.
\end{alignat}
These partitions correspond to the principal, diagonal and trivial embedding of $sl_2$ in $sl_3$, respectively.

The residual symmetries of the $\widehat{sl_N}$ WZW model constrained by \eqref{eq:ds-current-constraints} for the first two decompositions are the $\mathcal{W}_3$ algebra \cite{bershadsky-hidden-1989,forgacs-liouville-1989} and $\mathcal{W}_3^{(2)}$ Polyakov-Bershadsky algebra  \cite{bershadsky-conformal-1991}, respectively.
In addition to the Virasoro current, the former contains a spin three current, while the latter comes with two spin $3/2$ and a spin one current.
In the final decomposition, no positive radial weights appear so no constraints are imposed and we are still left with the full affine $\widehat{sl_2}$ current algebra.

More generally, we denote the reduced theory obtained from a general partition $\lambda$ by $\text{Toda}_\lambda$.
Its corresponding $\mathcal{W}_\lambda$ algebra contains a current for each $sl_2$ multiplet appearing in the decomposition of the adjoint of $sl_N$ \cite{bais_covariantly_1991,feher-hamiltonian-1992,deboer-relation-1994}.
As we will see in the following section, $\text{Toda}_\lambda$ can be obtained from six dimensions using the generalized conifold.\\

So far, we have been working with complex $sl_N$ models and their reductions, whereas the original AGT correspondence involves a real version of Toda theory.
To mediate this, the following relation is suggested for the principal embedding \cite{cordova-toda-2016}
$$
  \mathrm{complex \:Toda}(N,k,s)
    \;\;\Leftrightarrow \;\;
  \mathrm{real\: paraToda}(N,k,b)
    + \frac{\widehat{\mathfrak{su}}(k)_N}{\widehat{\mathfrak{u}}(1)^{k-1}}.
  $$
Here, $N-1$ gives the rank of the Toda theory.
The parameters $k$ and $s$ are coupling constants in the complex Toda theory.
On the right hand side, $k$ describes the conformal dimension $\Delta=1-1/k$ of the parafermions in paraToda.
The real Toda coupling is
$$
  b=\sqrt{\frac{k-is}{k+is}}.
$$

At $k=1$ the right hand side reduces to real Toda theory \cite{leclair_s-matrices_1993}.
Both real and complex generalized Toda theories can be obtained as Drinfeld-Sokolov reductions of $SL(N,\RR)$ or $SL(N,\CC)$ Chern-Simons theories.
Geometric quantization of the latter two theories yields identical Hilbert spaces \cite{dimofte-complex-2014} for $k=1$.
After reduction, the complex and real Toda theory therefore agree at this particular level.
Likewise, using a general $sl_2\subset sl_N$ embedding, complex and real $\text{Toda}_\lambda$ theory at $k=1$ are identified.

\section{Orbifold defects and the generalized conifold}\label{sec:defects-conifold}
In section \ref{ssec:codim2-defects} we briefly summarize the setup pertaining to the $\text{AGT}_\lambda$ correspondence.
This leads us to consider generalized conifolds, denoted by $\mathcal{K}^{k,m}$, whose geometry we review in section \ref{ssec:intersecting-d6-conifold}.

\subsection{Codimension two defects and their geometric realization}\label{ssec:codim2-defects}
We now consider the generalization of AGT that includes surface operators in the gauge theory partition function, which we refer to as $\text{AGT}_{\lambda}$.
Under the correspondence, the ramified instanton partition functions are mapped to conformal blocks of $\text{Toda}_{\lambda}$ theories \cite{wyllard_w-algebras_2011,wyllard_instanton_2011,tachikawa-walgebras-2011}.
Similarly, it is expected that one-loop determinants in the gauge theory map to three-point functions.
This has been checked for the case of a full surface operator \cite{nawata_givental_2014}.

A six-dimensional perspective on this correspondence is provided by including codimension two defects in the 6d $(2,0)$ theory.
These defects wrap $\Sigma$ and lie along a two-dimensional surface in the gauge theory.
Therefore, they represent a surface defect in the gauge theory, and change the theory on the Riemann surface.

There exists a natural class of codimension two defects that are labeled by partitions of $N$ \cite{gaiotto_n2_2012}, as we will discuss in more detail below.
The following table summarizes the M-theory background for the particular instance of $\text{AGT}_{\lambda}$ that we are interested in.
\begin{table}[H]
\centering
\begin{tabular}{lccccccccccc}
\toprule
& \multicolumn{4}{c}{$S^4$}&   \multicolumn{2}{c}{$\Sigma$}& \multicolumn{3}{c}{$\RR^3$}& \multicolumn{2}{c}{$\RR^2$}\\
\midrule
  & 0               & 1               & 2               & 3  & 4  & 5  & 6  & 7 & 8  & 9 &10  \\
\midrule
$N$ M5      & x              &x              & x              & x & x&x && & &&  \\
Defect & & &x  &x  & x& x &x &x & x & & \\
\bottomrule
\end{tabular}
\caption{Generalized AGT setup.}
\label{tab:desired-defect}
\end{table}
\noindent
For the moment, we zoom in on the region near the north pole of the $S^4$, where the geometry locally looks like $\CC^2$.
In this region, we consider the realization of the defect as a $\CC^2/\ZZ_m$ orbifold singularity that spans the 01910 directions of table \ref{tab:desired-defect} \cite{tachikawa-walgebras-2011,kanno-instanton-2011} (see also \cite{bullimore-superconformal-2015}).
Note that these codimension two defects are usually described using an additional set of intersecting M5 branes together with an orbifold singularity.
We want to emphasize here that we describe the defects using only the orbifold singularity.
This interpretation is also supported by mathematical results on the equivalence between ramified instantons and instantons on orbifolds.
See \cite{kanno-instanton-2011} and references therein.

This means that from the gauge theory perspective, i.e. the 0123 directions, that the geometry locally looks like $\CC/\ZZ_m\times \CC$.
A partition $\lambda$ is then naturally associated to the M5 branes
$$\lambda:\quad N=n_1+\ldots +n_m.$$
It specifies the number of M5 branes with a particular charge under the orbifold group.
Alternatively, when $\CC^2/\ZZ_m$ is thought of as a limit of an $m$-centered Taub-NUT space, it specifies how the M5 branes are distributed among the $m$ centers.
The M5 branes wrap the `cigars' in the second relative homology of $\mathrm{TN}_m$.
Upon reduction on the Taub-NUT circle fiber, the partition specifies how the $N$ D4 branes are distributed among the $m$ D6 branes.\\

Another generalization of the original AGT correspondence \cite{alday_affine_2010} that was already covered in the original derivation \cite{cordova-toda-2016} concerns instanton partition functions on $\CC^2/\ZZ_k$, an orbifold singularity that spans the 0123 directions of table \ref{tab:desired-defect} \cite{nishioka_para-liouville/toda_2011,wyllard_coset_2011}.
This generalization also naturally arises from a 3d-3d perspective since the $S^3/\ZZ_k$ in \eqref{eq:weyl-transf} is mapped to $S^4/\ZZ_k$ after the Weyl transformation.
The geometry near the north pole of this quotiented four-sphere is precisely $\CC^2/\ZZ_k$.\\

We now observe that there exists a simple (local) Calabi-Yau threefold that provides a particular realization of two ALE spaces $\CC^2/\ZZ_k$ and $\CC^2/\ZZ_m$, intersecting along a two-dimensional subspace.
This is the (partially resolved) generalized singular conifold
\begin{equation}\label{eq:generalized-conifold}
	\mathcal{K}^{k,m}:\quad xy=z^k w^m,
\end{equation}
where we identify $x$ or $y$ with the 01 directions, $z$ with the $23$ and $w$ with the 910 directions in the table \ref{tab:desired-defect}.
For earlier occurrences of this space, see \cite{uranga-brane-1999,dasgupta-brane-1999}.
More recently, it has also appeared in \cite{mcorist_relating_2011}.
Note that $\mathcal{K}^{1,m}$ reflects the $\text{AGT}_{\lambda}$ setup described above.
This will therefore be the geometry we focus on in the following, although we will also make a brief comment on general $k$ and $m$ in section \ref{ssec:general-K-km}.

We will use the generalized conifold as an approximation in the pole region of a squashed $S^4$ with defect included.
This is a considerable simplification to the full supergravity background that would be needed to preserve supersymmetry on the $S^4$ with defect.
One might worry that much information is lost by refraining from a similarly detailed and rigorous analysis as in \cite{cordova_complex_2013}.
However, we can still obtain a better understanding of the emergence of a general Nahm pole and the ensuing $\text{AGT}_{\lambda}$ correspondence.
This will be the main result of this paper.

\subsection{Intersecting D6s from the generalized conifold}\label{ssec:intersecting-d6-conifold}
In this section, we provide more detail on the geometry of our proposal.
First, we recall the relation between M-theory on a $\ZZ_k$ ALE space and $k$ D6 branes in IIA.
We then introduce the generalized conifold $\mathcal{K}^{k,m}$ and show how it effectively glues two such ALEs together into a single six-dimensional manifold, leading to two sets of $k$ and $m$ D6 branes upon reduction.\\

Consider the $\ZZ_k$ ALE space as a surface in $\CC^3$ described by
\begin{equation}
	xy = z^k.
\end{equation}
Equivalently, we can think of this four-dimensional space as a $\CC^2/\ZZ_k$ orbifold with
\begin{equation}
	1\in\ZZ_k: \quad
		(x,y)\in \CC^2 \to (e^{2\pi i/k}x, e^{-2\pi i/k}y).
\end{equation}
The latter makes it clear that the resulting space is singular.
In particular, we see that there is a $k$-fold angular deficit at the origin in the circle
\begin{equation}\label{eq:Zk-ALE-circle}
	C :=
		\left\{ ( e^{i\alpha} x, e^{-i\alpha} y) \,|\,
			e^{i\alpha} \in U(1)
		\right\}
	\simeq S^1.
\end{equation}
Reducing M-theory on the $\ZZ_k$ ALE along $C$ gives rise in IIA to $k$ D6 branes located at the origin and stretched along the transverse directions.
The generalized conifold $\mathcal{K}^{k,m}$ in equation \eqref{eq:generalized-conifold} describes a $\ZZ_k$ and $\ZZ_m$ ALE for fixed $w_0\neq0$ and $z_0\neq0$, respectively.
We will make these considerations more precise in the following.

In its most common form, which we will denote by $\mathcal{K}^{1,1}$, the standard conifold is a hypersurface in $\CC^4$ given by
\begin{equation}\label{eq:conifold-eqn-xyzw}
	xy = zw.
\end{equation}
The space $\mathcal{K}^{1,1}$ is a cone.
We denote its base by $T$, so that its metric is
\begin{equation}\label{eq:conifold-full-cone-metric}
	ds^2_{\mathcal{K}^{1,1}}
		= d\rho^2 + \rho^2 ds^2_{T}\,.
\end{equation}
It can easily be seen from \eqref{eq:conifold-eqn-xyzw} that $T$ is homeomorphic to $S^2\times S^3$.
For $\mathcal{K}^{1,1}$ to be K\"ahler, the base has to have the metric \cite{candelas-comments-1990}\footnote{
  Note that this reproduces the standard conifold metric upon redefining $\psi=(\psi'-\vphi_1-\vphi_2)/2$.
}
\begin{equation}\label{eq:conifold-base-metric}
\begin{split}
  ds^2_{T}
			&= \frac{4}{9}
    		\left( d\psi
    			+ \cos^2(\theta_1/2) d\vphi_1 + \cos^2(\theta_2/2) d\vphi_2 \right)^2 \\
    	&{}\qquad + \frac{1}{6} \left[
    		\left( d\theta_1^2 + \sin^2\theta_1 d\vphi_1^2 \right)
    		+ \left( d\theta_2^2 + \sin^2\theta_2 d\vphi_2^2 \right)
      \right].
\end{split}
\end{equation}
This describes two two-spheres, each with one unit of magnetic charge with respect to the shared Hopf fiber parametrized by $\psi$.
In other words, $T$ can be described by $SU(2)\times SU(2)/U(1)$.
The quotient by $U(1)$ serves to identify the Hopf fibers of the $SU(2)\simeq S^3$ factors.
More details can be found in appendix \ref{app:conifold}.

\begin{figure}
	\centering
	\includegraphics
		[width=.6\textwidth]
		{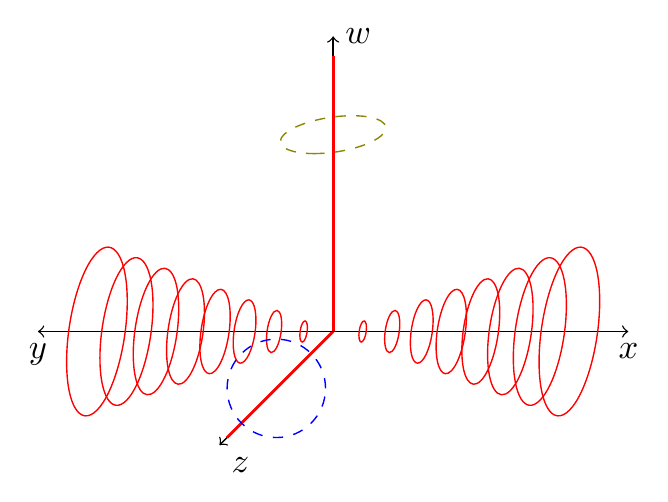}
	\caption{
		Each axis represents the modulus of a complex number, the surrounding circles denote its phase.
		We will reduce on the red (solid) circles, corresponding to the phase
    of $x$ and $y$, which shrink to a point at the $z$ and $w$ axes.
		}
	\label{fig:conifold-degenerate-fibers}
\end{figure}

\begin{figure}
	\centering
	\includegraphics
		[width=.6\textwidth]
		{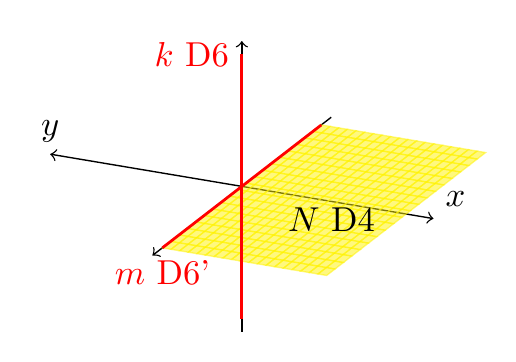}
	\caption{
	The resulting IIA brane content corresponding to $N$ M5 branes on the $w=0$ divisor $\mathfrak{D}_w$ of the generalized conifold $\mathcal{K}^{k,m}$ after reduction on the $xy$ circle.
	}
	\label{fig:conifold-D6s-figure}
\end{figure}

\vspace{\baselineskip}
Now we want to choose a circle which leads to intersecting D6 branes upon reduction to IIA.
Following the circle \eqref{eq:Zk-ALE-circle} in the ALE case, we are led to consider the action
\begin{equation}\label{eq:conifold-embedding-M-circle-u1-action}
  (x,y,z,w)
    \mapsto (e^{i\alpha}x, e^{-i\alpha}y, z,w),
	\qquad \alpha \in [0,2\pi).
\end{equation}
As can be seen from appendix \ref{app:conifold}, in terms of the Hopf coordinates in \eqref{eq:conifold-base-metric} describing the bulk of the base of the conifold, the circle is the orbit of
\begin{equation}\label{eq:conifold-su2-hopf-coords-u1-action}
  (\theta_1, \theta_2, \vphi_1, \vphi_2, \psi)
    \mapsto (\theta_1, \theta_2, \vphi_1 + \alpha,
							\vphi_2 +\alpha, \psi - \alpha).
\end{equation}
Thus the circle consists of equal $\theta_i$ orbits on the base two-spheres of $T$, together with a rotation in the Hopf fiber.
At $\theta_i=0$ or $\pi$ these Hopf coordinates are no longer valid and the circle described by \eqref{eq:conifold-su2-hopf-coords-u1-action} can shrink to a point.
In terms of the embedding $\CC^4$ coordinates, these loci are hypersurfaces $z=0$ and $w=0$, as we can see in \eqref{eq:conifold-embedding-M-circle-u1-action}.
This is illustrated in figure \ref{fig:conifold-degenerate-fibers}.
Reduction of M-theory along the circle generated by \eqref{eq:conifold-embedding-M-circle-u1-action} leads to two D6-branes stretched along the $z$ and $w$ directions, as illustrated in figure \ref{fig:conifold-D6s-figure}:
\begin{equation*}
	\text{ D6 from } 	x=y=z=0 \text{ along } w, \quad
	\text{ D6' from } 	x=y=w=0 \text{ along } z.
\end{equation*}

\vspace{\baselineskip}
Now let us look at the $w=0$ divisor, which we denote by $\mathfrak{D}_w$.
Setting $w=0$ in the conifold equation \eqref{eq:conifold-eqn-xyzw} implies that $x=0$ or $y=0$.
These are two branches meeting along the $z$ axis.
We will choose the latter one, so that the metric \eqref{eq:conifold-full-cone-metric} restricts to
\begin{equation}\label{eq:conifold-divisor-base-metric}
	ds^2_{\mathfrak{D}_w}
		= d\rho^2
			+ \frac{\rho^2}{6}\left(d\theta_2^2
        + \sin^2\theta_2 d\vphi_2^2 \right)
      + \frac{4\rho^2}{9} \left( d\psi
        + \cos^2 (\theta_2 /2) d\vphi_2 \right)^2.
\end{equation}
This is a radially fibered $S^3$ with a particular squashing.
Note that it preserves $SU(2)\times U(1)$ isometries.
At $\rho=1$ it is parametrized by (see appendix \ref{app:conifold})
\begin{equation}
\begin{split}\label{eq:conifold-divisor-hopf-glued-coords}
  &x = \cos \frac{\theta_2}{2} e^{i(\psi +\vphi_2)}, \\
  &z = \sin \frac{\theta_2}{2} e^{i\psi }.
\end{split}
\end{equation}
In these coordinates, the action \eqref{eq:conifold-embedding-M-circle-u1-action} whose orbit defines the M-theory circle is
\begin{equation}\label{eq:conifold-divisor-M-circle-u1-action}
  (\theta_2, \vphi_2, \psi)
    \to
  (\theta_2, \vphi_2 + \alpha, \psi).
\end{equation}
We see that the corresponding circles are just the equal $\theta_2$ circles of the second $SU(2)$ factor of the conifold, sitting at the north pole $\theta_1=0$ of the first $SU(2)$ factor.

Where does this circle shrink?
Again, we have to be careful about the range of our coordinates.
At the north pole $\theta_2=0$, the action \eqref{eq:conifold-embedding-M-circle-u1-action} shifts to the Hopf fiber,
\begin{equation}
	\theta_2 = 0: \quad
		(x,z) = (e^{i\beta},0), \quad
		\beta \to \beta + \alpha.
\end{equation}
That is a circle of finite size unless $\rho=0$.
In contrast, the orbit of \eqref{eq:conifold-embedding-M-circle-u1-action} shrinks to a point at the south pole $\theta_2=\pi$ for all $\rho$,
\begin{equation}
	\theta_2 = \pi: \quad
		(x,z) = (0, e^{i\delta}), \quad
		\delta \to \delta.
\end{equation}
Therefore, from the perspective of the divisor, the D6' brane stretches along $z$ at $x=y=0$ and the D6 brane is pointlike at $x=y=z=0$.\\

To obtain $m$ D6 and D6' branes, the M-theory circle should shrink with an $m$-fold angular deficit.
We can achieve this by quotienting the action \eqref{eq:conifold-embedding-M-circle-u1-action} by $\ZZ_m\subset U(1)$.
We denote the resulting generalized conifold by $\mathcal{K}^{m,m}$.\footnote{
Note that the labels on $\mathcal{K}^{m,n}$ are unrelated to the labels $(p,q)$ that are sometimes used to describe possible base spaces of the conifold.
}
It is given by
\begin{equation}\label{eq:singular-mm-conifold}
	xy = z^m w^m.
\end{equation}
We are mainly interested in the $w=0$ divisor $\mathfrak{D}_w$ of this space.
From the point of view of the divisor $\mathfrak{D}_w$, an $m$-fold angular deficit stretches along the $z$ axis.
Reducing to IIA leads to $m$ D6' branes that stretch along $z$ and are located at $x=w=0$.

A similar analysis for $z=0$, $w\neq0$ leads to a $m$-fold angular defect along $w$ at $x=y=z=0$.
This defect is pointlike in $\mathfrak{D}_w$, intersecting only at $x=y=w=z=0$.
Upon reduction to IIA, it leads to $m$ D6 branes.
The conifold point at the origin corresponds to the location where the two orbifold singularities intersect.\\

The generalized conifold $\mathcal{K}^{k,m}$ is described by equation \eqref{eq:generalized-conifold},
\begin{equation*}
	xy = z^k w^m.
\end{equation*}
It can be obtained by partially resolving the singularity along the $w$ axis of \eqref{eq:singular-mm-conifold}.
In a IIA frame, such a resolution corresponds to moving out $(m-k)$ D6 branes to infinity along the $x$ or $y$ axis.
The resulting geometry is similar to that of $\mathcal{K}^{m,m}$, except that it has a $k$-fold angular deficit intersecting at the origin with an $m$-fold one.
Consequently, reducing to IIA produces $k$ D6 branes and $m$ D6' branes.

By resolving all the way to $k=1$, $\mathfrak{D}_w$ is equivalent to the $\CC/\ZZ_m\times\CC$ background studied in \cite{kanno-instanton-2011}.
In terms of the coordinates \eqref{eq:conifold-divisor-hopf-glued-coords}, it is described by
\begin{align}\label{eq:conifold-divisor-k-hopf-glued-coords}
	&x = \cos \frac{\theta_2}{2} e^{i(\psi +\vphi_2/m)}, \\
	&z = \sin \frac{\theta_2}{2} e^{i\psi }.
\end{align}
The metric \eqref{eq:conifold-divisor-base-metric} then becomes
\begin{equation}\label{eq:conifold-divisor-kk-metric}
	ds^2_{\mathfrak{D}_w}
		= d\rho^2
			+ \frac{\rho^2}{6}\left(d\theta_2^2
        + \frac{1}{m^2} \sin^2\theta_2 d\vphi_2^2 \right)
      + \frac{4\rho^2}{9} \left( d\psi
        + \frac{1}{m} \cos^2 (\theta_2 /2) d\vphi_2^2 \right)^2.
\end{equation}
In this case, the sphere isometries are broken to $U(1)\times U(1)$.
The M-theory circle shrinks with a $\ZZ_m$ angular deficit along the $z$ axis.
Note that the three-spheres at fixed radius are in fact also squashed.
This leads to an additional `squashing' singularity at the origin, corresponding to the conifold point.\\

Finally, we should comment on how the M5 branes are placed in this geometry.
The $\mathfrak{D}_w$ divisor has two components ending on the $x=y=w=0$ defect, and we can place the M5s together along either one.
This setup preserves supersymmetry since the divisor is holomorphic.
See for instance \cite{ouyang-holomorphic-2004} for a similar setup in IIB.

Note that the M5 branes are in a sense fractional: a brane along the $x$ axis needs to pair up with a brane along the $y$ axis to be able to move off the defect.
Upon reduction, these fractional M5 branes correspond to D4 branes that end on the D6' branes, as we illustrate in figure \ref{fig:conifold-D6s-figure}.

As we will show in the next section, three of the scalars on the D4 branes will obtain a Nahm pole boundary condition dictated by how they are partitioned among the $m$ D6' branes.
On the other hand, the flux coming from the $k$ D6 branes gives rise to the Chern-Simons coupling in the 3d-3d frame.
Thus, both sets of D6 branes play distinct but crucial roles in our construction.

\section{\texorpdfstring{$\text{Toda}_{\lambda}$}{Toda(lambda)} theory from generalized conifolds}\label{sec:toda-from-conifold}
In this section, we outline a derivation of the $\text{AGT}_{\lambda}$ correspondence in the spirit of Córdova-Jafferis \cite{cordova-toda-2016}.
We will see that our proposal also sheds some light on the derivation of the original AGT correspondence.
The reason for this is that the surface operator associated with the trivial partition $N=N$ is decoupled from the field theory.
In our description, this is reflected by the fact that for $m=1$ there is no orbifold singularity.
Thus, we can view the original AGT correspondence as a special case of the $\text{AGT}_{\lambda}$ correspondence.
This will be discussed first.

Moving on to the general $\text{AGT}_\lambda$ correspondence, a crucial role is played by the generalized conifolds $\mathcal{K}^{1,m}$.
In the presence of our defect, the pole region of a squashed $S^4$ can be identified with an appropriate divisor in $\mathcal{K}^{1,m}$.
In this limit, the chirality and amount of the $S^4$ Killing spinors agree with those of the divisor in $\mathcal{K}^{1,m}$, which we take as further evidence for our proposal.

It should be noted that we use the (generalized) conifold merely as a technical simplification.
We expect a general set of supergravity backgrounds exists that allows for a squashed $S^4/\ZZ_k$ with defect included.
However, since the supersymmetry analysis is particularly easy for the conifold, we specialize to the parameter values it dictates.
Several subtleties that arise are expected to be resolved in the general set of supergravity backgrounds.

\subsection{Compatibility of \texorpdfstring{$\mathcal{K}^{1,1}$}{K1,1} with Córdova-Jafferis}
\label{ssec:K-11-compatibility}
As explained in the previous section, the standard conifold $\mathcal{K}^{1,1}$ produces a D6 and D6' brane if we reduce to type IIA.
In our setup, the D4 branes end properly on the D6' brane.
The latter is not present in the original derivation \cite{cordova-toda-2016} where only the D6 brane is considered.
The main reason for this discrepancy is that we choose a different circle fiber in \eqref{eq:conifold-embedding-M-circle-u1-action}.
In contrast to the Hopf fiber of the three-sphere, which only shrinks at the poles of the $S^4$, our circle degenerates along the entire $z$-plane in the pole region.

To check that we can still build on the results of \cite{cordova_complex_2013} we note that the D6 brane in \cite{cordova-toda-2016} translates to a flux for the graviphoton field in the 3d-3d frame, which is ultimately responsible for a non-zero Chern-Simons level in three dimensions.\footnote{See also \cite{dijkgraaf_supersymmetric_2008,itzhaki_i-brane_2006} for related discussions.}
This feature is not lost in our construction, since reduction on the conifold still produces a similar D6 brane located at the north pole.

Our brane configuration is illustrated in the following table.
\begin{table}[H]
\centering
\begin{tabular}{lccccccccccc}
\toprule
&$\RR$ & \multicolumn{2}{c}{$\RR^2$}&   \multicolumn{2}{c}{$\Sigma$}& \multicolumn{3}{c}{$\RR^3$}& \multicolumn{2}{c}{$\RR^2$}\\
\midrule
                 & 1               & 2               & 3  & 4  & 5  & 6  & 7 & 8  & 9 &10  \\
\midrule
$N$ D4                    &$\vdash$              & x              & x & x&x && & &&  \\
 D6'            &    &      x        & x & x&x &x&x &x &&\\
  D6           &    &             &  & x&x &x&x &x &x&x\\
\bottomrule
\end{tabular}
\caption{Type IIA setup in the pole region after reduction on the circle of $\mathcal{K}^{1,1}$.}
\label{tab:specialnahmpole}
\end{table}
\noindent
By the analysis of \cite{diaconescu_d-branes_1997,gaiotto_supersymmetric_2009}, the D4-D6' system leads to a (principal) Nahm pole boundary condition on three of the D4 scalars, denoted by $X_i$ in section \ref{ssec:cj-review}.
This provides a reinterpretation of the results in \cite{cordova-toda-2016}, where it is claimed that the D6 brane is responsible for the Nahm pole.

We will now turn to consistency checks of our identification of the conifold as an approximation of the supergravity background relevant to the AGT correspondence.
From the Nahm pole onwards, the original analysis of \cite{cordova-toda-2016} then goes through.
Namely, after Weyl rescaling to the 3d-3d frame, the Nahm pole translates into the reduction of the WZW model to Toda theory as reviewed in section \ref{sec:review}.

\paragraph{Geometry}
Recall that the metric on the $w=0$ divisor is given by \eqref{eq:conifold-divisor-base-metric},
\begin{equation*}
	ds^2_{\mathfrak{D}_w}
		= d\rho^2
			+ \frac{\rho^2}{6}\left(d\theta_2^2
        + \sin^2\theta_2 d\vphi_2^2 \right)
      + \frac{4\rho^2}{9} \left( d\psi
        + \cos^2 \theta_2 /2 d\vphi_2 \right)^2.
\end{equation*}
This is the pole region of the metric of a squashed four-sphere,
\begin{equation}\label{eq:squashed-s3-induced-metric-hopf-coordinates}
  ds^2 =d\sigma^2 + \left(\frac{f(\sigma)^2\ell^2}{4}\left(d\theta^2 + \sin^2(\theta) d\vphi^2 \right)
          +f(\sigma)^2 \left( d\psi + \cos^2(\theta/2) d\vphi \right)^2\right).
\end{equation}
The original derivation \cite{cordova-toda-2016} of the AGT correspondence considers the class of geometries \eqref{eq:squashed-s3-induced-metric-hopf-coordinates} for general $\ell$ and any $f(\sigma)$ that vanishes linearly near $\sigma=0,\pi$.
As we can see, the divisor of the conifold imposes the values $\ell=\tfrac{3}{\sqrt{6}}$ and $f(\sigma)=\tfrac{2}{3}\sigma + \OOO{\sigma^2}$.

Note that the conifold singularity translates to a squashing singularity of the metric \eqref{eq:squashed-s3-induced-metric-hopf-coordinates} for these particular choices of $f(\sigma)$ and $\ell$.
We will not be too concerned about these curvature singularities.
As argued in \cite{cordova-toda-2016}, the $(2,0)$ theory cannot couple to curvature scalars of dimension four or higher such as $R_{\mu\nu}R^{\mu\nu}$.
In principle, it could couple to the Ricci scalar, but this can be resolved by an appropriate choice of the function $f(\sigma)$.

For the singular conifold, which dictates $f(\sigma)$ and $\ell$ as above, the Ricci scalar singularity is present.
However, we expect the general set of supergravity backgrounds to contain solutions that exclude singularities in the Ricci scalar.
The curvature singularity should be merely an artifact of the parameter values imposed by the conifold.

\paragraph{Supersymmetries}
Here, we will show that the amount and chirality of supercharges preserved by the M5 brane on the conifold divisor match with what one expects for the 6d $(2,0)$ theory in the AGT setup.
Subsequently, we will relate them to the supercharges in the 3d-3d frame.

It is well known that an M5 brane wrapped on a holomorphic divisor inside a Calabi-Yau three-fold has at most $(0,4)$ supersymmetry in the remaining two dimensions \cite{maldacena_black_1997,minasian_calabi-yau_1999}.
However, since the latter lie along the Riemann surface $\Sigma$, on which the theory is topologically twisted, only two Killing spinors survive.\footnote{
This follows from the fact that the four supercharges form two doublets under the $SU(2)$ R-symmetry whose $U(1)\subset SU(2)$ subgroup is used to perform the twist.
}
Since the Killing spinors are chiral from both a six-dimensional and a two-dimensional perspective, they must be chiral in four dimensions as well,
\begin{equation*}
\xi^\text{chiral}_{6d}=\xi^\text{chiral}_{2d}\otimes \xi^\text{chiral}_{4d}.
\end{equation*}

Now let us turn to the usual AGT setup.
A 4d $\mathcal{N}=2$ theory on a squashed $S^4$ with $U(1)\times U(1)$ isometries has an $SU(2)_R$ doublet of Killing spinors \cite{hama-seibergwitten-2012,pestun_localization_2014},
\begin{align}\label{eq:hh-killingspinors}
\begin{split}
\varepsilon^1 &=(\xi_1,\bar{\xi}_1)=e^{\frac{1}{2}i(\phi_1+\phi_2)}\left(e^{-i\frac{\theta}{2}}\sin(\tfrac{\sigma}{2}),-e^{i\frac{\theta}{2}}\sin(\tfrac{\sigma}{2}),i e^{-i\frac{\theta}{2}}\cos(\tfrac{\sigma}{2}),-ie^{i\frac{\theta}{2}}\cos(\tfrac{\sigma}{2})\right)\\
\varepsilon^2 &=(\xi_2,\bar{\xi}_2)=e^{-\frac{1}{2}i(\phi_1+\phi_2)}\left(e^{-i\frac{\theta}{2}}\sin(\tfrac{\sigma}{2}),e^{i\frac{\theta}{2}}\sin(\tfrac{\sigma}{2}),-i e^{-i\frac{\theta}{2}}\cos(\tfrac{\sigma}{2}),-ie^{i\frac{\theta}{2}}\cos(\tfrac{\sigma}{2})\right).
\end{split}
\end{align}
Near the north pole these reduce, up to a local Lorentz and $SU(2)_R$ gauge transformation, to the $\Omega$ background
\begin{equation}\label{eq:killingspinors-Omegabg}
\bar{\xi}^{\dot{\alpha}}_A=\delta^{\dot{\alpha}}_A\;,\qquad \xi_{\alpha A}=-\frac{1}{2}v_m(\sigma^{m})_{\alpha\dot{\alpha}}\bar{\xi}^{\dot{\alpha}}_A.
\end{equation}
Here, $v_m$ is a Killing vector that generates a linear combination of the $U(1)^2$ isometry of the $\Omega$ background.
It descends from the $U(1)^2$ isometry of the squashed sphere.
Since $v_m$ vanishes linearly with $\sigma$ one sees that the Killing spinors are indeed chiral to zeroth order in $\sigma$.
Hence, the amount and chirality of the supercharges preserved by the divisor of the conifold are consistent with the ordinary AGT setup.

As noted in \cite{cordova-toda-2016}, after Weyl rescaling to the 3d-3d frame, in a suitable R-symmetry gauge, the Killing spinors in \eqref{eq:hh-killingspinors} become independent of $\sigma$.
This is required to make contact with the Killing spinors in the 3d-3d correspondence \cite{cordova_complex_2013} which are independent of $\sigma$ due to the topological twist on $M_3$.

A final subtlety is to be mentioned here.
In the above, we have made contact with the Killing spinors corresponding to the $
\mathcal{N}=2$ theory on a squashed $S^4$ which preserves $U(1)\times U(1)$ isometries.
However, the derivation of the 3d-3d correspondence in \cite{cordova_complex_2013} makes use of a squashed sphere with $SU(2)\times U(1)$ isometries.
As was first observed in \cite{imamura_n2_2012} and then properly understood in \cite{closset_geometry_2014}, the three-dimensional supersymmetric partition function is in fact insensitive to these extra symmetries.
This should provide a justification for the proposed relation between the partition functions evaluated in the 4d-2d and 3d-3d frame.

\paragraph{Twist}
Another perspective on the equivalence of the preserved supersymmetries in the conifold case and the AGT setup lies in their relation to topological twists.
The worldvolume theory on the M5 branes is automatically topologically twisted, since it wraps a K\"ahler cycle inside a Calabi-Yau threefold \cite{bershadsky_d-branes_1996}.\footnote{The normal bundle to the divisor is its canonical bundle. Then, the two scalars corresponding to transverse movement inside $\mathcal{K}^{1,1}$ become holomorphic two-forms on the divisor \cite{maldacena_black_1997,minasian_calabi-yau_1999}.}
In terms of groups, the R-symmetry is broken by the setup to $U(2)$. The $U(1)$ R-symmetry given by the embedding
$$U(1)\subset U(2)\subset SO(4)\subset SO(5)$$
is used to twist the $U(1)\subset U(2)$ holonomy on the divisor (see e.g. appendix A in \cite{gadde_fivebranes_2013}).

This feature is also reflected in the standard AGT setup.
Indeed, at zeroth order in $\sigma$, the Killing spinors \eqref{eq:killingspinors-Omegabg} precisely reflect the ordinary (Donaldson-Witten) topological twist: the $SU(2)_R$ index is identified with the dotted spinors index.
The twist implemented by the conifold is a special version of this twist when the holonomy is reduced to $U(2)$.

\subsection{\texorpdfstring{$\mathcal{K}^{1,m}$}{K1,m} and \texorpdfstring{$\text{AGT}_{\lambda}$}{AGT(lambda)}}
\label{ssec:K-1m-agt}
We now want to explain the relevance of the generalized conifold for the $\text{AGT}_\lambda$ correspondence.
First, a partition is associated to the divisor that specifies the charges of the fractional M5 branes in the orbifold background
\begin{equation}\label{eq:partition}
\lambda: \qquad N=n_1+\ldots+n_m.
\end{equation}
After reduction on the circle fiber, this partition encodes the number $n_i$ of D4 branes ending on the $i^{\mathrm{th}}$ D6' brane.
\begin{table}[H]
\centering
\begin{tabular}{lccccccccccc}
\toprule
&$\RR$ & \multicolumn{2}{c}{$\RR^2$}&   \multicolumn{2}{c}{$\Sigma$}& \multicolumn{3}{c}{$\RR^3$}& \multicolumn{2}{c}{$\RR^2$}\\
\midrule
                 & 1               & 2               & 3  & 4  & 5  & 6  & 7 & 8  & 9 &10  \\
\midrule
$N$ D4                    &$\vdash$              & x              & x & x&x && & &&  \\
 $m$ D6'            &    &      x        & x & x&x &x&x &x &&\\
  D6           &    &             &  & x&x &x&x &x &x&x\\
\bottomrule
\end{tabular}
\caption{Type IIA setup in the pole region after reduction on the circle of $\mathcal{K}^{1,m}$.}
\label{tab:generalnahmpole}
\end{table}
\noindent
This imposes a Nahm pole on the three $X^a$ D4 worldvolume scalars in terms of the $sl_2\subset sl_N$ embedding associated to $\lambda$.
In the 3d-3d frame, the D6 brane is solely reflected as a graviphoton flux which leads to a $k=1$ Chern-Simons level \cite{cordova_complex_2013}.

In the M-theory frame, Weyl rescaling from 4d-2d to 3d-3d results in an asymptotically hyperbolic three-manifold $M_3$ corresponding to the directions $145$ in table \ref{tab:generalnahmpole}.
The half-line along the $1$ direction is stretched to a line with an asymptotic boundary.
Since the D6' branes are located at the edge of this half-line, the Nahm pole they induce becomes a constraint at the asymptotic boundary of $M_3$ in the Weyl rescaled frame.

The bulk of $M_3$ is therefore unaffected by the presence of the additional D6' branes introduced by the generalized conifold.
The analysis of \cite{cordova_complex_2013}, which obtains complex Chern-Simons theory from reduction of the M5 worldvolume theory, should then still hold in the bulk of $M_3$.
We thus claim that only the constraint on the boundary behavior of the Chern-Simons theory is different.

To be precise, the partition of the $N$ D4 branes on $m$ D6' branes translates in the 3d-3d frame to a block diagonal form of the connection at the boundary.
For example,
\begin{equation}
  \lambda: \quad 3 = 2 +1
    \qquad\longleftrightarrow\qquad
  \mathcal{A} =
  \begin{pmatrix}
    \ast  & \ast  & \\
    \ast  & \ast  & \\
          &       & \ast
  \end{pmatrix}.
\end{equation}
The Nahm pole then maps to a constraint in this block diagonal form,
\begin{equation}
  \mathcal{A} = d\rho + e^\rho \LL^+ dz + \cdots,
    \qquad
  \LL^+ =
  \begin{pmatrix}
    0   & 1   & \\
    0   & 0   & \\
        &     & 0
  \end{pmatrix}.
\end{equation}
As we have outlined in section \ref{ssec:dsreduction}, such a constraint precisely reduces a $SL(N)$ WZW model to the Toda theory associated to the partition $\lambda$.
Recalling the equivalence between complex and real $SL(N)$ Chern-Simons at level $k=1$ \cite{dimofte-complex-2014}, this leads to a derivation of the $\text{AGT}_\lambda$ correspondence.

\paragraph{Geometry}
The $w=0$ divisor of $\mathcal{K}^{1,m}$ is equivalent to $\CC/\ZZ_m\times \CC$, as we showed in section \ref{ssec:intersecting-d6-conifold}.
The non-trivial $\Omega$ background that is manifested by the squashing of a radially fibered three-sphere, just as in the $k=m=1$ case, is also visible there.
The squashed three-sphere in this geometry preserves $U(1)\times U(1)$ isometries since the base of the Hopf fibration is now orbifolded.
This shows that the divisor in $\mathcal{K}^{1,m}$ reproduces the setup in which the $\text{AGT}_\lambda$ correspondence was studied \cite{kanno-instanton-2011}.\footnote{The superconformal index of the 6d $(2,0)$ theory in the presence of these orbifold singularities was computed in \cite{bullimore-superconformal-2015}.}

\paragraph{Supersymmetries}
The generalized conifolds preserve the same amount of supersymmetry as the $m=1$ conifold.
This is particularly clear from the IIA perspective, where instead of a single D6 and D6' brane, we now have one D6 intersecting with $m$ coincident D6' branes.

Likewise, the general AGT setup concerns a squashed $S^4$ with $U(1)\times U(1)$ isometries,
$$
  t^2
    + \frac{\left|z\right|^2}{\ell^2}
    + \frac{\left|x\right|^2}{\tilde{\ell}^2}
  =1
$$
It is clear that including our defect at $x=0$ does not break these isometries any further.
Therefore, just as in our conifold construction, including such defects in the general AGT setup does not break any additional supersymmetry.

Furthermore, the divisor of the generalized conifold is still K\"ahler, so only chiral supersymmetries survive.
This agrees with the chirality of the Killing spinors of an $S^4$ in the pole region.

\subsection{General \texorpdfstring{$k$}{k} and \texorpdfstring{$m$}{m}}
\label{ssec:general-K-km}
Finally, we comment on a conjecture arising from the general conifold $\mathcal{K}^{k,m}$.
Here, we expect to obtain $SL(N,\CC)$ Chern-Simons theory at level $k$ together with a boundary condition determined by the partition
$$\lambda: \;N=n_1+\ldots+n_m.$$
According to this partition one should obtain a quantum Drinfeld-Sokolov reduction of complex Toda theory.
To arrive at a duality with a real paraToda theory, in the spirit of \cite{cordova-toda-2016}, one could naively ask if
$$
  \mathrm{Complex \:Toda_{\lambda}}(n,k,s)
    \;\;\stackrel{?}{\Leftrightarrow} \;\;
  \mathrm{real\: paraToda}_{\lambda}(n,k,b)
    +\frac{\widehat{\mathfrak{su}}(k)_n}{\widehat{\mathfrak{u}}(1)^{k-1}}.$$
However, such a statement requires one to understand how parafermions couple to generalized Toda theories.
We are not aware of the existence of any such constructions.
An obvious first step would be to figure out how parafermions could couple to affine subsectors.

\section{Conclusions and outlook}
\label{sec:concl}
We have argued that the derivation of the AGT correspondence proposed in \cite{cordova-toda-2016} can be understood by replacing the north pole region of the $S^4$,
where the excitations of the four-dimensional gauge theory are localized,
with a holomorphic divisor inside the singular conifold $\mathcal{K}^{1,1}$.
This interpretation has two main virtues.
Firstly, it provides a clear perspective on the origin of the Nahm pole.
Secondly, the generalized conifolds $\mathcal{K}^{1,m}$ allow us to outline a generalization of the derivation of the original AGT correspondence, which we denote by $\text{AGT}_\lambda$, involving the inclusion of surface operators on the gauge theory side and a generalization of Toda theory to $\text{Toda}_\lambda$ on the two-dimensional side.
We used an equivalent description of these surface operators as orbifold defects, as advocated in \cite{tachikawa-walgebras-2011,kanno-instanton-2011}.
\\

Let us now turn to the possible pitfalls of our analysis and their potential resolutions.
First of all, we make a number of assumptions and simplifications due to a lack of a full supergravity background and supersymmetry equations of the worldvolume theory in the presence our additional defects.
Even in the original case, a full supergravity background and supersymmetry equations have only been written down for the 3d-3d frame \cite{cordova_complex_2013}.
For a precise understanding of the origin of the Nahm pole, one should furthermore obtain the supergravity background and supersymmetry equations of the 4d-2d frame.
In the presence of our additional defects, such a background should include a geometry which resembles a conifold near its pole regions.
Transforming the supersymmetry equations to the 3d-3d frame should then give rise to the Drinfeld-Sokolov boundary conditions on the Chern-Simons connection.

An obstruction to performing this transformation is that these equations can only be written down in a five-dimensional setting, since the Lagrangian formulation of the 6d $A_{N-1}$ theory is unknown.
In other words, one cannot directly transform the supersymmetry equations leading to a Nahm pole in 4d-2d to the 3d-3d equivalent which should give the correct Chern-Simons boundary conditions.

Furthermore, the derivation of the 3d-3d correspondence \cite{cordova_complex_2013} is in a sense only concerned with the bulk of the Chern-Simons theory.
Even there, ghosts appear, which gauge fix the noncompact part of the gauge group at finite $S^3_\ell/\ZZ_k$ size.
On an $M_3$ with boundary, one should similarly impose boundary conditions on these ghosts, which are related to the boundary constraints of the Chern-Simons connection.
It would be interesting to see if such ghost terms can be obtained in the 3d-3d frame, if they correspond to the proper Drinfeld-Sokolov constraints in the setting we propose, and how they translate to the 4d-2d Nahm pole setting.

At level $k=1$, using the equality between the Hilbert spaces of complex and real $SL(N)$ Chern-Simons theory, the complex $\text{Toda}_\lambda$ theory corresponds to a real $\text{Toda}_\lambda$ theory.
For higher $k$, it would be interesting to understand the equivalent of the parafermions that were necessary to make contact with real Toda in the original derivation \cite{cordova-toda-2016}.
\\

In the main body of this paper, we have not touched upon the relation between (a limit of) the superconformal index of the 6d $(2,0)$ theory of type $A_{N-1}$ and vacuum characters of $\mathcal{W}_N$ algebras discovered in \cite{beem-w-2015}.
This relation was also derived in \cite{cordova-toda-2016} following similar arguments to their derivation of the AGT correspondence.
The geometry relevant to the superconformal index, $S^5\times S^1$, can be Weyl rescaled to $S^3 \times EAdS_3$.
One can understand this by thinking of the $S^5$ as an $S^3$ fibration over a disc, where the $S^3$ shrinks at the boundary of the disc.
The Weyl rescaling stretches the radial direction of the disk to infinite length and produces the $EAdS_3$ geometry.

The boundary conditions on the Chern-Simons connection are again argued to be of Drinfeld-Sokolov type.
The D6 brane that arises from reduction on the Hopf fiber wraps the boundary circle of the disc and the $S^1$.

As in the derivation of the AGT correspondence, this D6 brane does not have the correct codimensions for a D4 brane to end on it, and for its scalars to acquire a Nahm pole.
Again, we claim that the analogous identification of the divisor of a conifold in the $S^5$ reproduces the D6 brane and additionally produces the D6' on which the D4s can end.
This construction generalizes to the inclusion of codimension two defects as orbifold singularities, as studied in \cite{bullimore-superconformal-2015}.
This leads to a derivation of the conjecture, appearing before in \cite{beem-w-2015}, that the vacuum character of a general $\mathcal{W}_\lambda$ algebra is equal to the 6d superconformal index.\\

Further possibly interesting directions of research include the following.
In the 3d-3d frame, the additional defects we introduced affect the boundary conditions of Chern-Simons theory.
On the other hand, they also have two directions along the three-dimensional $\mathcal{N}=2$ theory $T[M_3]$.
It would be interesting to interpret the role these defects play on this side of the correspondence.

We can also include other types of defects.
Codimension two defects that are pointlike on the Riemann surface translate to operator insertions in the Toda theory.
They are similarly labeled by a partition of $N$, which we can associate to the choice of a (possibly semi-degenerate) Toda primary.
In six dimensions, these defects wrap an $S^4$ that maps under Weyl rescaling to an $S^3$ times the radial direction of $M_3$.
One can then couple such a codimension two defect to the five-dimensional Yang-Mills theory.
Reducing to $M_3$ should produce a Wilson line in complex Chern-Simons theory.

Finally, the central charge of generalized Toda theories is known for any embedding, see for example \cite{deboer-relation-1994}.
It would be interesting to reproduce this central charge from six dimensions.
This has been done for principal Toda in \cite{alday_liouville/toda_2010} by equivariantly integrating the anomaly eight-form over the $\RR^4$ $\Omega$ background.
Following the geometric description of the codimension two defects, one could integrate a suitable generalization of the anomaly polynomial on the orbifolded $\CC\times\CC/\ZZ_m$ $\Omega$ background.
Reproducing the generalized Toda central charge from such a computation would provide a convincing check on the validity of a geometric description of the codimension two defects.

\section*{Acknowledgments}
It is a pleasure to thank Jan de Boer, Clay Córdova and Erik Verlinde for useful discussions and encouragement, and Yuji Tachikawa for correspondence on the geometric realization of codimension two defects.
This work is partially supported by the $\Delta$-ITP consortium and the Foundation for Fundamental Research on Matter (FOM), which are funded by the Dutch Ministry of Education, Culture and Science (OCW).

\appendix

\section{The conifold}\label{app:conifold}
Let us review some facts on the conifold.
We mainly follow \cite{candelas-comments-1990} but choose slightly different coordinates in places.
The conifold $\mathcal{K}^{1,1}$ is a hypersurface in $\CC^4$ defined by
\begin{equation}\label{eq:app-conifold-eqn-xyzw}
	zw = xy.
\end{equation}
This equation defines a six-dimensional cone.
By intersecting $\mathcal{K}^{1,1}$ with a seven-sphere of radius $r$, we can study its base, which we denote by $T$.
The base $T$ is topologically equivalent to $S^2\times S^3$ and can be conveniently parametrized in the following way,
\begin{equation}
	\label{eq:app-conifold-base-u2-equations}
	Z :=
	\frac{1}{r}
		\begin{pmatrix}
			z & x \\
			y & w
		\end{pmatrix}
	\qquad
	T: \quad
  \det Z = 0, \quad
  \Tr Z^\dagger Z = 1.
\end{equation}
We can write down the most general solution to these equations by taking a particular solution $Z_0$ and conjugating it with a pair $(L,R)$ of $SU(2)$ matrices,
\begin{equation}\label{eq:app-conifold-su2-base-solution}
  Z = L Z_0 R^\dagger, \qquad
  Z_0 =
    \begin{pmatrix}
      0 &  1 \\
      0 &  0
    \end{pmatrix},
    \quad
    L,R \in SU(2).
\end{equation}
Each $SU(2)$ factor can be described using two complex coordinates,
\begin{equation}\label{eq:app-conifold-su2-factors-param}
\begin{split}
  L &:=
  \begin{pmatrix}
    a   &   -\bar{b} \\
    b   &   \bar{a}
  \end{pmatrix}
    \in SU(2),
  \quad
  (a,b)\in \CC^2,
  \quad
  |a|^2 + |b|^2 = 1, \\
  R &:=
  \begin{pmatrix}
    k   &   -\bar{l} \\
    l   &   \bar{k}
  \end{pmatrix}
    \in SU(2),
  \quad
  (k,l)\in \CC^2,
  \quad
  |k|^2 + |l|^2 = 1.
\end{split}
\end{equation}
Now introduce Hopf coordinates on each $SU(2)\simeq S^3$,
\begin{equation}
    \label{eq:app-conifold-su2-proper-hopf-coords}
\begin{split}
  &a = \cos(\theta_1/2) e^{i(\psi_1+\vphi_1)},
    \qquad
    k = \cos(\theta_2/2) e^{i(\psi_2+\vphi_2)}, \\
  &b = \sin(\theta_1/2) e^{i\psi_1},
    \qquad\qquad
    l = \sin(\theta_2/2) e^{i\psi_2},
\end{split}
\end{equation}
Note that the parametrization of $T$ in \eqref{eq:app-conifold-su2-base-solution} is overcomplete.
Two pairs of $SU(2)$ matrices $(L,R)$ describe the same solution if and only if they are related by the $U(1)$ action
\begin{equation}
  (L,R) \mapsto (L \Theta, R \Theta^\dagger),
		\qquad
	\Theta =
		\begin{pmatrix}
			e^{i\theta} & 0 \\
			0 & e^{-i\theta}
		\end{pmatrix}
		\in U(1) \subset SU(2).
\end{equation}
This degeneracy should be quotiented out of the $SU(2)\times SU(2)$ parametrization.
The $U(1)$ acts on the $S^3$ coordinates by
\begin{equation}\label{eq:app-conifold-quotient-from-c4}
  (a,b)\to (e^{i\theta}a, e^{i\theta}b), \quad
  (k,l)\to (e^{-i\theta}k, e^{-i\theta}l).
\end{equation}
The resulting $SU(2)\times SU(2)/U(1)$ quotient is the conifold.
Indeed, the invariant coordinates under this $U(1)$ action correspond to the ones used in \eqref{eq:app-conifold-base-u2-equations}.
Setting $r=1$,
\begin{equation}
  x = ak, \quad
  y = -bl, \quad
  z = -al, \quad
  w = bk.
\end{equation}
They are related by the defining equation \eqref{eq:app-conifold-eqn-xyzw} of the conifold.
In terms of the Hopf coordinates \eqref{eq:app-conifold-su2-proper-hopf-coords}, the $U(1)$ quotient \eqref{eq:app-conifold-quotient-from-c4} joins the two Hopf fiber coordinates in the invariant combination $\psi:= \psi_1 + \psi_2$.
Then $T$ is parametrized by
\begin{subequations}\label{eq:app-conifold-hopf-glued-coords}
\begin{align}
  x &= \cos \frac{\theta_1}{2} \cos \frac{\theta_2}{2}
        e^{i(\psi + \vphi_1 + \vphi_2)}, \\
  y &= - \sin \frac{\theta_1}{2} \sin \frac{\theta_2}{2}
        e^{i\psi}, \\
  z &= - \cos \frac{\theta_1}{2} \sin \frac{\theta_2}{2}
        e^{i(\psi + \vphi_1)}, \\
  w &= \sin \frac{\theta_1}{2} \cos \frac{\theta_2}{2}
        e^{i(\psi + \vphi_2)}.
\end{align}
\end{subequations}
Demanding that $\mathcal{K}^{1,1}$ is Kähler implies that the metric on $T$ is given by \cite{candelas-comments-1990}
\begin{align}
  ds^2_{T}
    &= \frac{2}{3} \Tr \left( dZ^\dagger dZ \right)
      - \frac{2}{9} \left|\Tr \left( Z^\dagger dZ \right)\right|^2 \\
    &=
			\frac{4}{9}
    		\left( d\psi
    			+ \cos^2(\theta_1/2) d\vphi_1 + \cos^2(\theta_2/2) d\vphi_2 \right)^2 \\
    	&{}\qquad + \frac{1}{6} \left[
    		\left( d\theta_1^2 + \sin^2\theta_1 d\vphi_1^2 \right)
    		+ \left( d\theta_2^2 + \sin^2\theta_2 d\vphi_2^2 \right)
      \right]. \nonumber
\end{align}
This is the metric we wrote down in \eqref{eq:conifold-base-metric}.
It describes two three-spheres with a shared Hopf fiber.
If we think of this fibration as an electromagnetic $U(1)$ bundle, both spheres feel one unit of magnetic charge.
Note that this metric is equivalent to the usual one under the coordinate redefinition $\psi = (\psi' - \vphi_1 - \vphi_2)/2$.

\paragraph{The divisor}
In the main text, we make extensive use of the $w=0$ divisor of the conifold.
Setting $w$ to zero in \eqref{eq:app-conifold-eqn-xyzw} implies that either $x$ or $y$ vanishes.
In terms of the coordinates in \eqref{eq:app-conifold-hopf-glued-coords}, these choices corresponds to setting either $\theta_1=0$ or $\theta_2=\pi$.
Thus we are at the north (or south) pole of one of the $S^2$ base factors of $T$.
The remaining sphere, together with the fiber, now describes an ordinary $S^3$.
Setting $\theta_1=0$, the parametrization in \eqref{eq:app-conifold-hopf-glued-coords} reduces to Hopf coordinates
\begin{equation}
\begin{split}\label{eq:app-conifold-divisor-hopf-glued-coords}
  &x = \cos \frac{\theta_2}{2} e^{i(\psi +\vphi_2)}, \\
  &z = \sin \frac{\theta_2}{2} e^{i\psi }.
\end{split}
\end{equation}

\bibliographystyle{JHEP}
\bibliography{complete-bibliography}

\end{document}